\documentclass[sigchi, natbib=false]{acmart}
\usepackage{amsmath,amssymb,amsfonts}
\usepackage{algorithmic}
\usepackage{graphicx}
\usepackage[utf8]{inputenc}
\usepackage{textcomp}
\usepackage{xcolor}
\usepackage[
    style=numeric-comp,
    backend=biber,
    natbib=true,
    url=true, 
    doi=false,
    eprint=false,
    mincitenames=1,
    maxcitenames=1,
    maxnames=1,
    giveninits=true,
    sorting=none,
    bibencoding=utf8
]{biblatex}
\usepackage{subcaption}				% two figures next to each other (example: figure 3a), figure 3b)

\usepackage{listings, color}			% for source code
\usepackage{url}           			% format URLs
\usepackage{textcomp}				% for special symbols like '°'
\usepackage{booktabs}
\usepackage{xcolor}
\usepackage{enumitem}

\usepackage{xspace}

\usepackage{amsmath} 
\usepackage{amsthm}
\usepackage{amssymb}

\usepackage[]{todonotes}

\usepackage{pgfplots}
\pgfplotsset{
/pgfplots/ybar legend/.style={
/pgfplots/legend image code/.code={%
\draw[##1,/tikz/.cd,bar width=3pt,yshift=-0.2em,bar shift=0pt]
plot coordinates {(0cm,0.8em)};},
},
}

\usepackage[capitalise]{cleveref}	% for clever references

\newlist{requirements}{enumerate}{1}
\setlist[requirements]{label*=(\arabic*), ref=\arabic*}

\crefname{requirementsi}{requirement}{requirements}
\Crefname{requirementsi}{Req.}{Req.}

\newcommand{\squeezeup}{\vspace{-2.5mm}}

\newcommand{\tuple}[5]{$\langle$\textit{#1}, \textit{#2}, \textit{#3}$\rangle$}

\graphicspath{{./img/}}

\def\BibTeX{{\rm B\kern-.05em{\sc i\kern-.025em b}\kern-.08m 
    T\kern-.1667em\lower.7ex\hbox{E}\kern-.125emX}}
\begin{document}

\acmConference[IOT’19]{9th International Conference on the Internet of Things}{October
22--25}{Bilbao, Spain}

\title{Blockchain-based Data Provenance for the \\ Internet of Things} % \thanks{Identify applicable funding agency here. If none, delete this.}}

\author{Marten Sigwart}
\affiliation{%
\institution{TU Wien, Vienna, Austria}
%\city{Vienna}
%\country{Austria}
}
\email{m.sigwart@infosys.tuwien.ac.at}

\author{Michael Borkowski}
\affiliation{%
\institution{TU Wien, Vienna, Austria}
%\city{Vienna}
%\country{Austria}
}
\email{m.borkowski@infosys.tuwien.ac.at}

\author{Marco Peise}
\affiliation{%
\institution{TU Berlin, Berlin, Germany}
%\city{Berlin}
%\country{Germany}}
}
\email{mp@ise.tu-berlin.de}

\author{Stefan Schulte}
\affiliation{%
\institution{TU Wien, Vienna, Austria}
%\city{Vienna}
%\country{Austria}
}
\email{s.schulte@infosys.tuwien.ac.at}

\author{Stefan Tai}
\affiliation{%
\institution{TU Berlin, Berlin, Germany}
%\city{Berlin}
%\country{Germany}
}
\email{st@ise.tu-berlin.de}

\acmYear{2019}
\copyrightyear{2019}

\begin{abstract}
As more and more applications and services depend on data collected and provided by Internet of Things (IoT) devices, it is of importance that such data can be trusted. Data provenance solutions together with blockchain technology are one way to make data more trustworthy. However, current solutions do not address the heterogeneous nature of IoT applications and their data. 

In this work, we identify functional and non-functional requirements for a generic IoT data provenance framework, and conceptualise the framework as a layered architecture. Using a proof-of-concept implementation based on Ethereum smart contracts, data provenance can be realised for a wide range of IoT use cases. Benefits of a generic framework include simplified adoption and a more rapid implementation of data provenance for the IoT.

\end{abstract}

\maketitle
\renewcommand{\shortauthors}{Sigwart et al.}


\section{Introduction}
\label{sec:intro}
The Internet of Things (IoT) is transforming many areas of our everyday lives. IoT technologies such as GPS, RFID-based identification, and low-resource computing platforms like the Raspberry Pi already play important roles in various domains, e.g., mobility, logistics, healthcare, and retail~\cite{li2015internet}. Since data collected by these technologies find application in an increasing number of use cases, ensuring the trustworthiness of such data is of high importance~\cite{aman2017secure}.
%\begin{figure*}
%	\centering
%	\includegraphics[width=(0.70\textwidth)]{vaccine-supply-chain-iot}
%	\caption{IoT-enhanced Vaccine Supply Chain}
%	\label{fig:vaccine-supply-chain-iot}
%\end{figure*}

Data provenance systems are one way to ensure trustworthiness in the IoT~\cite{bertino2014data}. These systems can provide information about the origin and evolution of data such as the various stages of data creation and data modifications, who initiated them, and when and how they took place~\cite{freire2008provenance}. In short, a data provenance system tracks who has created, updated, deleted, and in some cases read particular data points~\cite{freire2008provenance}. In order to trust the information provided by a provenance system, it is essential that the provenance system stores information in a tamper-proof and replicable way~\cite{hasan2009preventing}.

Traditionally, in distributed settings, participants must either trust each other or an independent third-party to store data in a tamper-proof way. With the advent of blockchain technologies, the requirement of such trust in a central authority is eliminated. Instead, a decentralised network can be established, acting as a distributed, tamper-proof ledger~\cite{christidis2016blockchains}. Thus, leveraging blockchain technology in a data provenance solution for the IoT is a promising choice~\cite{christidis2016blockchains}.

The IoT is characterised by a multitude of use cases and potential application areas~\cite{li2015internet}. An IoT data provenance solution has to account for this diversity~\cite{olufowobi2017data}. However, approaches aiming at scenario-agnostic blockchain-based data provenance solutions for the IoT offer so far no concrete software solutions~\cite{polyzos2017blockchain, baracaldo2017securing}, and more concrete solutions to data provenance in the IoT only focus on specific application areas, for instance, supply chains~\cite{toyoda2017novel, westerkamp2018blockchain, wu2017distributed}, health monitoring systems~\cite{griggs2018healthcare}, or digital forensics~\cite{cebe2018block4forensic}.

Hence, we propose a generic blockchain-based data provenance framework for the IoT that can be applied to a variety of use cases. The advantages of a generic framework are the easier adoption of provenance concepts by new use cases and the interoperability of applications that use the framework. The contributions of this work are as follows:

\begin{itemize}[leftmargin=*]
	\item We define functional and non-functional requirements for a generic IoT data provenance framework.
	\item We conceptualize and implement an IoT data provenance framework consisting of smart contracts using a generic data model to provide provenance functionality for a wide range of IoT use cases.
	\item We present an evaluation of the framework with regard to the defined requirements using a proof-of-concept implementation with Ethereum smart contracts.
\end{itemize}

%\begin{itemize}
%	\item A framework consisting of smart contracts which use a generic data provenance model to provide provenance functionality to a wide range of IoT use cases.
%	\item A review of recent developments in blockchain technologies with regards to scalability, suitable for an IoT data provenance system.
%	\item A proof-of-concept implementation based on Ethereum smart contracts and Plasma sidechains, presenting and evaluating the framework.
%\end{itemize}

Following, we briefly explain underlying concepts and provide exemplary use case scenarios~(\cref{sec:background}), define the requirements~(\cref{sec:requirements}), and explain the concepts, architecture, and proof-of-concept implementation of our IoT data provenance framework~(\cref{sec:framework}). In~\cref{sec:eval}, we evaluate the framework with regard to the defined requirements. \Cref{sec:related} provides an overview of the related work. Finally, \cref{sec:conclusion} concludes the paper.

\section{Background \& Motivation}
\label{sec:background}

Data provenance, sometimes also known as lineage or pedigree~\cite{simmhan2005survey}, identifies the derivation history of data~\cite{freire2008provenance}. While originally used for works of art, data provenance is now relevant in a wide range of use cases, since data provenance mechanisms help to establish a certain level of trust in data by providing information about its creation, access, and transfer~\cite{freire2008provenance}. Provided provenance data is secured, forgery, alteration or repudiation of data can be prevented~\cite{hasan2009preventing}. Accordingly, data provenance solutions can help to establish trust in data in the IoT~\cite{aman2017secure}. In the following, we describe two exemplary use cases for data provenance in the IoT.

\paragraph{Vaccine Supply Chains}
\label{sec:usecase1}
Immunisation programmes depend on functional, end-to-end supply chains~\cite{comes2018cold}. In all phases of the supply chain---from procurement to last-mile distribution---it is of significant importance that vaccines remain in a temperature range of \mbox{around 2--8 °C}. Otherwise, vaccines lose their effectiveness. An unbroken, temperature-controlled link from producer to consumer is also referred to as the cold chain~\cite{comes2018cold}. In IoT-enabled cold chains, technologies such as RFID, GPS and sensing technologies~\cite{li2015internet} track temperatures, locations and other conditions of the vaccines. This provenance data helps to establish confidence in the quality of vaccines and exposes any weak links along the supply chain. As mentioned in \cref{sec:intro}, an important requirement for provenance systems is that recorded provenance information is replicable and tamper-proof. Recent scandals\footnote{\url{https://www.securingindustry.com/pharmaceuticals/massive-fake-vaccine-racket-busted-in-indonesia/s40/a2849/\#.WrylEtNuaL4}} of vaccine counterfeiting have confirmed the urgency of these requirements in vaccine supply chain systems. Hence, an IoT-powered data provenance system based on blockchain technology could provide benefits in vaccine supply chain scenarios. In particular, it enables the backtracking of errors in case of breakage of the cold chain, and it helps to build and to keep trust in immunisation programmes by preventing counterfeit vaccines from entering the supply chain.

%\cref{fig:vaccine-supply-chain-iot} shows an exemplary vaccine supply chain. As vaccines travel along the cold chain, IoT technologies such as RFID-based identification, GPS and sensing technologies~\cite{li2015internet} track temperatures, location and other conditions of the vaccines. This provenance data helps to establish confidence in the quality of vaccines and exposes any weak links along the supply chain.

%\paragraph{Health Monitoring Systems}
%\label{usecase2}
\textit{Health Monitoring Systems.} Health monitoring scenarios are another application area where tamper-proof data provenance records can provide additional trust in data~\cite{griggs2018healthcare}. In such scenarios, sensors monitor health conditions of patients such as a patient's heart rate, blood pressure, etc. Combined with a real-time analytics service, the sensor readings can be used to notify relatives and health professionals in case of medical emergencies, such as a heart attack. Ideally, the root cause of such a notification is verifiable, i.e., it is possible to trace back a notification to the individual sensor readings which caused it. Otherwise, if the circumstances that led to the emergency notification are unclear, the possibility that the notification was caused by a malicious attack tampering with the system cannot be excluded. Having tamper-proof provenance data for a notification, such as information about the individual sensor readings and analytics involved, ensures the reliability and accuracy of the system.

%\paragraph{Digital Forensics}
%\label{usecase3}
%\textit{Digital Forensics.} In digital forensics scenarios, for instance, in the case of accidents involving autonomous vehicles~\cite{cebe2018block4forensic}, sensors within and around the vehicle continuously record important data such as speed, road conditions, outside temperature, etc. while driving. Additionally, maintenance data of the vehicles is provided by garages. Ideally, in case of an accident, investigators can use such data to solve any disputes between the involved parties, such as drivers, insurance companies, garages, etc. Therefore, it is crucial that the integrity of the data is ensured and that the source of the data can be trusted. Yet again, data provenance records secured via blockchain technology can provide such benefits.
\section{Requirements}
\label{sec:requirements}
This section defines the functional and non-functional requirements of a generic data provenance framework for the IoT. Such a framework needs to record provenance information for addressable data points, i.e., data that has a unique ID~\cite{olufowobi2017data}. In accordance with~\cite{freire2008provenance} and~\cite{olufowobi2017data}, we derive the functional requirements~(\Crefrange{req:abstraction}{req:parallel}) of our framework. Further, in accordance with Hasan et al.~\cite{hasan2009preventing}, we define non-functional requirements that need to be fulfilled by a tamper-proof data provenance system for the IoT~(\Crefrange{req:integrity}{req:scalability}).

\begin{requirements}[leftmargin=*]
	\item \textit{Provenance Abstraction:} The framework needs to provide generic data provenance capturing, storing, and querying functionality which can be adopted by provenance use cases to map their specific requirements. \label{req:abstraction} %\vspace{1mm}

	\item \textit{High-level and Low-level Provenance:} Provenance records represent high-level as well as low-level data points. Low-level data points stem from low-level devices such as sensors. High-level data points do not have a single physical origin (such as a sensor reading) but represent more abstract concepts, e.g., some physical object in a supply chain or an analytics result based on multiple inputs.\label{req:high-level}%\vspace{1mm}

	\item \textit{Completeness:} A provenance record is complete if every relevant action which has ever been performed on a data point is gathered~\cite{hasan2009preventing}. Here, relevance implies that some actions can be neglected if they do not contribute to the provenance information of a particular data point.\label{req:complete}%\vspace{1mm}
	
	\item \textit{Creation of Lineage:} Provenance records for a data point can be created based on the last provenance record for that same data point. This enables the creation of lineage. For instance, the lineage of a data point representing a physical good travelling along a supply chain can be tracked by creating a new provenance record based on the old one at each critical step of the supply chain.\label{req:lineage}%\vspace{1mm}
	
	\item \textit{Derivation:} A provenance record entails references to the provenance records of the data points that led to its creation. For instance, a provenance record for an analytics result based on value readings from multiple sensors must not only contain information about the sensor values but must also be able to access provenance information of those same values, such as location and time of recording.\label{req:derivation}%\vspace{1mm}

	\item \textit{Provenance for Modifications of Data Points:} The framework enables the tracking of the modification history of a specific data point. For instance, if an analytics result runs through multiple stages of different calculations, the history of these calculations can be tracked.\label{req:modification}%\vspace{1mm}
	
	\item \textit{Parallel Provenance:} Multiple provenance records for the same data point can exist in parallel, e.g., one provenance trace might track the ownership of a data point (e.g., ownership of a physical good), while another provenance record is tracking the location of that same data point.\label{req:parallel}%\vspace{1mm}
	
	\item \textit{Integrity:}  Integrity mandates that provenance records cannot be manipulated or modified by an adversary in any way. This is crucial for establishing trust in the data. Without guaranteed integrity, clients can potentially repudiate provenance records.\label{req:integrity}%\vspace{1mm}
	
	\item \textit{Availability:} Data can be accessed when needed.\label{req:availability}%\vspace{1mm}
		
	\item \textit{Privacy:} Provenance records of IoT devices can contain sensitive data, e.g., in a health monitoring system. It is therefore vital to keep this data confidential, i.e., to prevent access by unauthorised entities. In addition, even when the data itself is kept confidential, malicious actors might still be able to create user profiles by identifying user-specific data patterns. Hence, traceability of provenance data needs to be prevented to ensure user privacy.\label{req:privacy}
	
	%Privacy usually can be distinguished into confidentiality and anonymity. On the one hand, provenance records of IoT devices might contain sensitive data, e.g., in a health monitoring system. It is therefore vital to keep this data confidential, i.e., access to this data from unauthorised entities needs to be prevented. On the other hand, even when the data itself is kept confidential, malicious actors might still be able to create user profiles by identifying user-specific patterns. As such, traceability of provenance data needs to be prevented to ensure the anonymity of users. In the following, we refer to both notions simply as privacy.\label{req:privacy}

	\item \textit{Scalability:} A provenance system is required to have a reasonable expenditure. Storing and accessing provenance information must have a low overhead. Especially, even though some IoT devices are resource-constrained, they must not be excluded from participating in the provenance framework. Further, applications in the IoT potentially deal with massive amounts of data and very frequent data updates which a provenance solution needs to account for.\label{req:scalability}%\vspace{1mm}

\end{requirements}
\section{IoT Data Provenance Framework}
\label{sec:framework}

\begin{figure}
	\centering
	\includegraphics[width=0.49\textwidth, height=4.2cm]{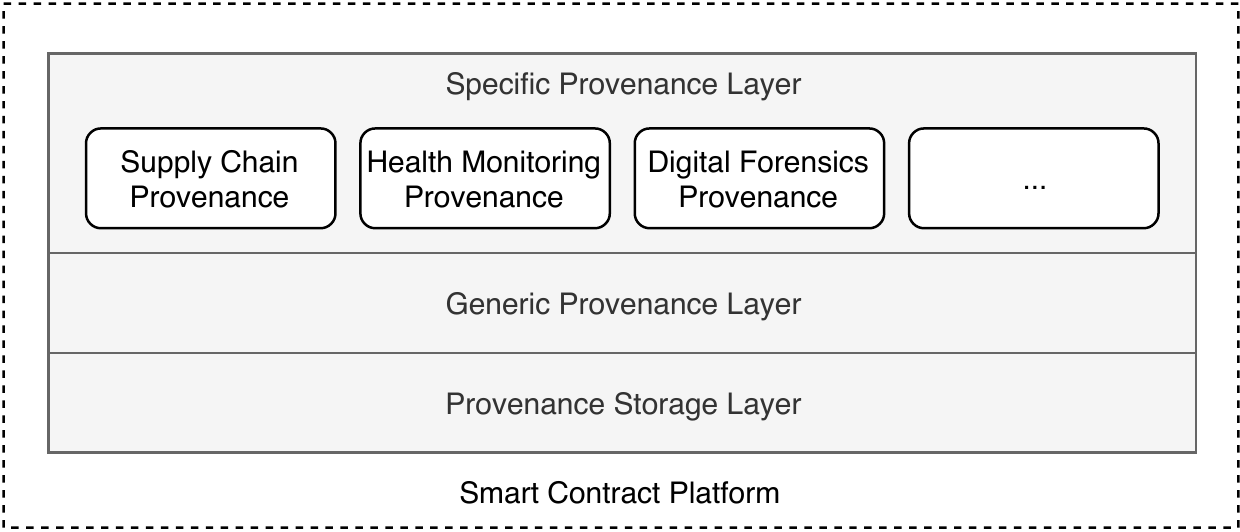}
	\caption{Data Provenance Framework Architecture}
	\squeezeup
	\label{fig:architecture}
\end{figure}

This section presents the proposed generic blockchain-based data provenance framework for the IoT. As mentioned in~\cref{sec:intro}, the benefits of a generic framework are interoperability between applications using the framework and the facilitated implementation of provenance concepts for new IoT use cases. \Cref{fig:architecture} displays the core architecture of the framework. It consists of three layers all embedded within a blockchain smart contract platform. Each layer represents a different level of abstraction with a diverse set of responsibilities within the framework. The storage layer is primarily concerned with low-level representation and storage of provenance data, the generic provenance layer provides general-purpose provenance functionality, while the specific provenance layer can be modified by use cases to fine-tune the framework to their specific requirements. 

A proof-of-concept implementation of the framework based on Ethereum smart contracts is available as open-source software at Github\footnote{\url{https://github.com/msigwart/iotprovenance}}.

\subsection{Data Model}
\label{sec:data-model}
Provenance information varies depending on the specific domain or application~\cite{olufowobi2017data}. Since the IoT domain is characterised by heterogeneous applications and a wide range of possible use cases~\cite{li2015internet}, there is a need for a generalised provenance model suitable for IoT applications. As the underlying data model, our framework utilises the data provenance model by Olufowobi et al.~\cite{olufowobi2017data} since their model is specifically designed to address IoT provenance data. The model defines a provenance record for some data point \textit{dp} as follows: 

\begin{displaymath}
\begin{aligned}
	\textit{prov} : \textit{dp} \mapsto & \langle \textit{addr}(\textit{dp}), \{\textit{prov}(\textit{idp}) | \forall\textit{idp} \in \textit{inputs}(\textit{dp})\}, \\ 
	& \textit{context}(\textit{dp})\rangle
\end{aligned}
\end{displaymath}
	
A provenance record \textit{prov}$($\textit{dp}$)$ for a data point \textit{dp} consists of a 3-tuple associating the address \textit{addr}$($\textit{dp}$)$ of \textit{dp}~(i.e., some kind of ID) with the set of provenance records of the data points that led to the creation of \textit{dp}~(i.e., \textit{inputs}$($\textit{dp}$)$), and a context~\textit{context}$($\textit{dp}$)$. The context denotes information of interest for provenance about the state of the IoT system, e.g., information about agents involved in the computation of the data point, time and location, or the execution context. This definition of provenance allows for the description of both creation and modification of data points. In the former case, the set of input provenance records contains an empty set. In the latter case, the set of input provenance records contains the provenance record of the data point before modification, i.e., \(\textit{prov}(\textit{dp}') = \langle \textit{addr}(\textit{dp}'), \{\textit{prov}(\textit{dp}), ...\}, \allowbreak\textit{context}(\textit{dp}')\rangle \).

The described data provenance model combined with blockchain technology builds the basis for our IoT data provenance framework. This way, the framework can not only represent provenance information for various IoT use cases, but also provides integrity guarantees for the recorded data.

\begin{lstlisting}[caption={Storage Contract (Excerpt)	}, label=lst:provenance-records, float]
contract Storage {
struct ProvenanceRecord {
    uint tokenId;
    uint[] inputProvenanceIds;
    string context;
    uint index;
}
mapping (uint -> ProvenanceRecord) records;
uint[] provenanceIndex;
function getProvenance(uint provId)
public view returns (uint tokenId, ...) {...}
...
function createProvenance(uint provId, ...) 
internal returns (uint index) {...}
...
}
\end{lstlisting}

\subsection{Storage Layer}
\label{sec:storage-layer}
This layer is responsible for the low-level storage of provenance records. It contains the generic representation for provenance records as defined by the data provenance model, as well as basic functionality to create, retrieve, update, and delete provenance records. Delete refers to the invalidation of provenance records, since truly deleting data from a blockchain is not possible. An invalidated provenance record cannot be used as input for subsequent provenance records.

\Cref{lst:provenance-records} displays an excerpt of the smart contract implementing this layer. The internal representation of provenance records (Lines 2--7) closely resembles the model discussed above with the field \textit{tokenId} representing the ID of a data point. However, not only are data points addressable, but also the provenance records themselves. Addressable provenance records allow the storage layer to manage provenance records as a mapping from provenance IDs to provenance records: $\textit{addr}(\textit{prov}) \mapsto \textit{prov}$ where the function \textit{addr}$($\textit{prov}$)$ represents the ID of a provenance record \textit{prov}. The mapping is implemented via the \textit{mapping} keyword~(Line 8). 

%The contract further defines a field \textit{provenanceIndex} implemented as an array containing all provenance IDs within the system~(Line~9). These IDs can be used for iteration over provenance records.
The contract exposes an API for creating, retrieving, updating, and deleting (i.e., invalidating) provenance records. Note that, while functionality for retrieving provenance records is exposed publicly via the \textit{public} keyword (Lines~10ff), functionality for creating, updating, and deleting records is protected via the \textit{internal} keyword (Lines~13ff). These functions cannot be accessed publicly, but are accessible from inheriting contracts such as the contract representing the generic provenance layer. Read-functions are publicly accessible since these functions do not alter the state of the contract.

\subsection{Generic Provenance Layer}
\label{sec:generic-provenance-layer}
The generic provenance layer's main purpose is to provide general-purpose provenance functionality on top of the storage layer, i.e., it provides features that are universally applicable for a wide range of provenance use cases. The generic provenance layer has the following responsibilities.

\paragraph{Ownership of Data Points}
\label{sec:design-ownership} 
While blockchain technology can guarantee the integrity of data provenance records once those records have entered the system, mechanisms need to be in place to ensure that the records that enter the system are correct. As a first step, we aim to prevent the creation of provenance records by arbitrary clients, i.e., if client $A$ generates some data point $\textit{dp}_0$, we want to make sure that only client $A$ (or any client authorised by client $A$) is able to create provenance records for $\textit{dp}_0$. Thus, we introduce the notion of ownership of data points. Each data point belongs to a specific client of the system, and only the owner or a client authorised by the owner can create provenance records for it. If an unauthorised client tries to create a provenance record for a data point, the system raises an error.%\hlr{of course this still does not mitigate the issue of authorised clients creating illegal provenance records...}

The notion of ownership is closely related to so-called tokens, i.e., smart contracts deployed on a blockchain that represent a kind of digital asset~\cite{tokens}. Our framework leverages tokens to introduce ownership of data points. Each data point is represented by a single token and a single token identifies exactly one data point. This one-to-one mapping allows us to identify the owner of a data point by identifying the owner of a particular token. 

Each token acts as an entry ticket to the provenance framework. To create provenance records for a particular data point, a client first has to become the owner or be approved by the owner of the corresponding token. The prototype uses the Ethereum token standard ERC721\footnote{\url{https://github.com/ethereum/EIPs/blob/master/EIPS/eip-721.md}} which defines a common interface for non-fungible assets, for instance, functions for transferring ownership. This has the advantage that our tokens (i.e., the data points) can be traded by any client implementing this standard, such as wallets or exchanges. By transferring ownership, data points can pass from owner to owner, leaving a trail of provenance records created by each owner along the way. This is useful for implementing provenance applications, e.g., in supply chain scenarios.% The generic provenance contract complies to the standard by inheriting from an existing ERC721 implementation provided by OpenZeppelin\footnote{\url{https://openzeppelin.org/}}.

\paragraph{Associating Provenance Records with Data Points}
\label{sec:datapoint-provenance-mapping}
The generic provenance layer further links together data points and their respective provenance records. The framework provides information about associated provenance records of specific data points. Within the generic provenance layer, this is achieved by using a mapping from a data point~ID to a set of associated provenance IDs:
\begin{displaymath}
	\textit{addr}(\textit{dp}) \mapsto \{\textit{addr}(\textit{prov}_{1}(\textit{dp})), \textit{addr}(\textit{prov}_{2}(\textit{dp})), ...\}
\end{displaymath}
All associated provenance records (\textit{prov}$_{1}$, \textit{prov}$_{2}$, etc.) represent parallel provenance traces for the same data point. Of those records, each one represents a completely independent trail of provenance information. For instance, one trail of provenance records might trace the temperature history of a physical good, while another one might trace its location.

\Cref{lst:associated-provenance} displays an excerpt of the smart contract implementing this layer and demonstrates the general workflow for creating new provenance records. First, the contract verifies that a given token (i.e., data point) exists~(Line~7) and that the token belongs to the sender of the message~(Line~8). After validating the input provenance records~(Line~9), the contract creates a new provenance ID~(Line~10), adds it to the list of associated provenance~(Line~11), and calls the storage contract's \textit{createProvenance} function~(Line~12). Note that the \textit{createProvenance} function of the generic contract is again internal, and thus not accessible publicly. This enables specific provenance contracts which implement concrete use cases to further adapt the functionality according to their needs. 

\begin{lstlisting}[caption={Generic Provenance Contract (Excerpt)}, label={lst:associated-provenance}, float]
contract GenericProvenance is ERC721, Storage {
...
mapping (uint => uint[]) internal associatedProv;
...
function createProvenance(uint tokenId, ...) 
internal returns (uint index) {
    require(exists(tokenId));
    require(ownerOf(tokenId) == msg.sender);
    checkValidProvenance(inputProvenance);
    uint provId = getProvId();
    addAssociatedProvenance(tokenId, provId);
    return super.createProvenance(provId, ...);
}
...
\end{lstlisting}

\subsection{Specific Provenance Layer}
\label{sec:specific-provenance-layer}
\begin{lstlisting}[caption={Specific Provenance Contract (Example)}, label={lst:specific-provenance}, float]
contract Specific is GenericProvenance {
function requestToken() 
public returns (uint tokenId) {
    uint tokenId = getTokenId();
    return super.mint(msg.sender, tokenId);
}
function createProvenance(uint tokenId, 
string location, uint temperature, ...) 
public returns (uint index) {
    string context = createContext(location, 
    temperature, ...);
    return super.createProvenance(provId, 
    context, ...);
}
...
\end{lstlisting}
Smart contracts within this layer utilise the functionality provided by the generic provenance layer, but control a subset of parameters by themselves. This way, use cases can customise the provenance model according to their needs, and control access to the functionality provided by the storage and generic provenance layer. \Cref{lst:specific-provenance} displays an excerpt of an exemplary smart contract implementing this layer. 

The provenance model can be customised by defining the \textit{context} parameter, so that it presents the provenance information needed for a specific use case scenario. For instance, in the case of vaccine supply chains, the context should contain temperature and location information about individual vaccines. Hence, a specific contract could define its own \textit{createProvenance} function that requires parameters like \textit{temperature}, and \textit{location}~(Lines~7ff) which are then combined to form the \textit{context} to be passed on to the \textit{createProvenance} function of the generic provenance contract~(Lines~10ff).

Access control happens on two levels. First, a specific contract defines which parts of the generic provenance layer's API are exposed. For instance, even though the generic provenance layer could permit updating or deleting (i.e., invalidating) provenance records, this could be unwanted behaviour in the specific use case at hand. In this case, the contract in the specific provenance layer simply ``hides'' the functionality, i.e., does not expose it publicly.

Second, access control is relevant for controlling the ownership of data points. Since data point ownership is the decisive factor with regard to who can create provenance records for which particular data points, contracts in the specific provenance layer are responsible for actually assigning ownership, i.e., which tokens get assigned to which clients. For simplicity, our prototype automatically assigns new tokens to requesting clients~(Lines~2ff). However, ultimately, the most suitable approach is dictated by the specific use case at hand. For instance, clients could purchase tokens. This would keep the framework publicly available while reducing the risk of spamming attacks since the acquisition of tokens incurs financial cost. Another alternative is a white list of authorised clients allowed to request new tokens. This way, the list controls exactly who participates in the provenance system. However, the question arises who is responsible for managing the white list of authorised clients.

%
%\begin{itemize}[leftmargin=*]
%	\item One possibility is for clients to purchase tokens. This has the advantage that the framework is publicly available without the risk of spamming attacks since the acquisition of tokens incurs financial cost.
%	%Essentially, the token acquisition costs need to be low enough for honest users to be willing and able to participate, but high enough to discourage spammers. Of course, such an approach cannot prevent malicious actors with sufficient purchasing power.
%	\item In another approach, the layer manages a white list of authorised clients allowed to request new tokens. This way, the list controls exactly who participates in the provenance system. However, the question arises who is responsible for managing the white list of authorised clients.
%	%\item Yet another approach is a list of pending requests. A client sends a transaction requesting tokens. An administrator gets notified about the new request and decides whether to accept or deny the request by assigning or not assigning corresponding tokens. While this approach allows a more fine-grained control over whose requests get accepted and whose requests get denied, it bears the risk of a high degree of centralisation due to the administrator having full control over distributing tokens.
%\end{itemize}
%\input{sections/04implementation.tex}
\section{Evaluation}
\label{sec:eval}
We evaluate the framework with regard to the functional and non-functional requirements defined in~\cref{sec:requirements}. The use cases defined in \cref{sec:background} act as basis for the evaluation of \Crefrange{req:abstraction}{req:complete}. In addition, we reason about the fulfilment of \Crefrange{req:lineage}{req:parallel} in an exemplary fashion applying the use case of vaccine supply chains. However, the scenario of vaccine supply chains is merely used to provide a more descriptive analysis. The information specific to the vaccine supply chain can be substituted with data reflecting any other use case. While the presented framework is blockchain-agnostic, the non-functional requirements~(\Crefrange{req:integrity}{req:scalability}) are evaluated using the proof-of-concept implementation. Experiments are performed on the public Ethereum test networks Rinkeby\footnote{\url{https://rinkeby.etherscan.io/}} and Ropsten\footnote{\url{https://ropsten.etherscan.io/}}. Rinkeby and Ropsten are chosen as test networks since their average block times most closely resemble the block times of the main Ethereum network.% Rinkeby and Ropsten use Proof of Authority (PoA) and Proof of Work~(PoW), respectively.

%Ropsten is the test network that most closely resembles the production environment of the main Ethereum network since it uses PoW as consensus algorithm. A potential problem using Ropsten is that the network can be subject to spamming attacks~\cite{ropsten-attacks}, since everyone can mine ETH at will and there is no real incentive to secure the system other than altruistic reasons (ETH in a test network has no economic value). Rinkeby on the other hand, guarantees a more stable network since it uses PoA as consensus algorithm. Rinkeby has an average block time of about 15~seconds which is similar to the average block times of the main Ethereum network~(see \cref{fig:chart-blocktimes}). Other possible test networks are Kovan\footnote{\url{https://kovan.etherscan.io/}} and Sokol\footnote{\url{https://sokol-explorer.poa.network/}}. However, Kovan and Sokol were not considered for the experiments because their average block times differ substantially from the main network ($\sim4$ and $\sim5$ second block times, respectively).

\paragraph{Provenance Abstraction~(\Cref{req:abstraction})} The presented framework consists of multiple abstraction layers. The storage layer is responsible for low-level storage of provenance records while the generic provenance layer extends the storage layer's functionality with generic provenance features and enhanced access control. The generic provenance layer can be extended by use case-specific smart contracts controlling certain parameters of the application, e.g., by assigning ownership of tokens, and/or by exposing or hiding parts of the generic layer's API. For instance, in the scenario of vaccine supply chains, a specific provenance smart contract might give an entity such as the World Health Organisation complete control over who is able to acquire tokens, e.g., only trusted vaccine manufacturers. In the use case of health monitoring systems, multiple approved manufacturers might have control over assigning ownership of data point IDs. Hence, we conclude that the framework acts as a base abstraction which can be extended to fulfil the needs of specific provenance use cases. Thus, we regard \Cref{req:abstraction} as fulfilled. %In the use case of digital forensics, each newly purchased vehicle might automatically get assigned a new token representing that vehicle. Besides creating provenance records itself, the vehicle could give temporary permission to garages or other entities to create provenance records on its behalf, e.g., to register the date and specifics of an inspection.

\textit{High-level and Low-level Provenance~(\Cref{req:high-level}).} The framework identifies each data point by its corresponding token. As long as a unique token (i.e., ID) gets assigned to a data point, provenance information for that data point can be recorded. Hence, the framework does not pose any restrictions on the nature of the data point. It is possible to record provenance information for high-level as well as low-level data points. The \emph{context} parameter of a provenance record can be used to add any kind of information the user desires. In the use case of vaccine supply chains, provenance information could not only be collected for the vaccines themselves, but also for sensor readings along the supply chain, e.g., temperature readings which document an uninterrupted cold chain. Within health monitoring systems, the patients themselves might be represented by data points. The patients' provenance traces are then augmented by provenance information from low-level data points deriving from medical devices. Hence, we regard \Cref{req:high-level} as fulfilled. %Similarly, in the use case of autonomous vehicles, the vehicles themselves are represented by data points and their provenance records get augmented by data points deriving from low-level devices, such as a speedometer or outside temperature sensors. 

\textit{Completeness~(\Cref{req:complete}).} The broad definition of the \emph{context} parameter allows the collection of all information necessary for complete provenance tracking. The definition of completeness is largely dependent on the specific provenance use case. In the example of vaccine supply chains, information required to prove the origin of vaccines and an uninterrupted cold chain can be recorded. Regarding health monitoring systems, we are able to record all information necessary to provide reliable root cause information for medical emergencies. Therefore, \Cref{req:complete} can also be regarded as fulfilled. %With regard to autonomous vehicles, the system is able to record all information necessary to resolve any disputes between parties involved in an accident. 

\textit{Creation of Lineage~(\Cref{req:lineage}).}
As an example, we assume that the manufacturer \textit{SaferVaccines Inc.} produces a new vaccine \textit{vacc$_{1}$}. The freshly produced vaccine is packaged with an RFID tag and assigned with a unique ID. A completely new provenance record for the vaccine is created, such as $\textit{prov}(\textit{vacc}_{1})$$=$\allowbreak\tuple{addr$($vacc$_{1})$}{$\emptyset$}{\tuple{agent$=$operator$_{1}$@Safer\-Vaccines\-Inc}{time$=$\allowbreak5am}{...}}~~. Vaccine~\textit{vacc$_{1}$} is loaded onto an aircraft~\textit{air$_{1}$}. An RFID reader at the factory gate scans the vaccine and registers a new provenance record of the vaccine leaving the factory, such as \textit{prov$($vacc${1}$}$)'$$=$\tuple{addr$($vacc$_{1})$}{$\{$prov$($vacc$_{1})\}$}{\tuple{agent$=$\allowbreak rfid$_{1}$\-@Safer\-Vac\-cines\-Inc}{time$=$5am}{...}}~~. Shortly after, an RFID reader at the aircraft's entrance registers the vaccine entering the aircraft, e.g., \textit{prov$($vacc$_{1})''$$=$}\allowbreak\tuple{addr$($vacc$_{1})$}{$\{$prov$($vacc$_{1})'\}$}{\tuple{agent$=$rfid$_{1}$\-@air$_{1}$}{time$=$5am}{...}}~~. The provenance records \textit{prov$($vacc$_{1})$}, \textit{prov$($vacc$_{1})'$}, and \allowbreak\textit{prov$($vacc$_{1})''$} represent the lineage of the vaccine. This shows exemplarily how the framework fulfils \Cref{req:lineage}.

\textit{Derivation~(\Cref{req:derivation}).}
Continuing the example from \Cref{req:lineage}, inside the aircraft, sensors constantly monitor the temperature. There are three readings from a temperature sensor: \textit{dp}$_{1}=38^\circ$F, \textit{dp}$_{2}=45^\circ$F, \textit{dp}$_{3}=40^\circ$F. The provenance records are given as follows:\par
	
$\textit{prov}(\textit{dp}_{1})$$=$\tuple{addr$(\textit{dp}_{1})$}{$\emptyset$}{\tuple{agent$=$sensor}{time$=$7am}{...}}~~\par
		
$\textit{prov}(\textit{dp}_{2})$$=$\tuple{addr$(\textit{dp}_{2})$}{$\emptyset$}{\tuple{agent$=$sensor}{time$=$8am}{...}}~~\par
		
$\textit{prov}(\textit{dp}_{3})$$=$\tuple{addr$(\textit{dp}_{3})$}{$\emptyset$}{\tuple{agent$=$sensor}{time$=$9am}{...}}~~\par
		
\noindent A fourth data point $\textit{dp}_{4}$ is created by a software agent calculating the average temperature, i.e., $\textit{dp}_{4}=(\textit{dp}_{1}+\textit{dp}_{2}+\textit{dp}_{3})/3 = 41^\circ$F. This data point's provenance record is defined as $\textit{prov}(\textit{dp}_{4})$$=$\tuple{addr$(\textit{dp}_{4})$}{$\{\textit{prov}(\textit{dp}_{1}),\textit{prov}(\textit{dp}_{2}),\textit{prov}(\textit{dp}_{3})\}$}{\allowbreak\tuple{agent$=$ave\-rager\-@air$_{1}$}{time$=$10am}{...}}~~. Hence, the framework allows for the creation of provenance records for data points deriving from multiple other data points~(\Cref{req:derivation}).

\textit{Provenance for Modifications of Data Points~(\Cref{req:modification}).}
In a further step, the calculated average temperature $\textit{dp}_{4}$ is converted into a different unit: $\textit{dp}_{4}' = 5^\circ$C~(i.e., from Fahrenheit to Celsius). The resulting provenance record looks like $\textit{prov}(\textit{dp}_{4}')$$=$\tuple{addr$(\textit{dp}_{4}')$}{$\{\textit{prov}(\textit{dp}_{4})\}$}{\tuple{agent$=$con\-ver\-ter@air$_{1}$}{time$=$11am}{...}}~~. Thus, the framework is also able to support the modification of data points~(\Cref{req:modification}).

\textit{Parallel Provenance~(\Cref{req:parallel}).}
Besides measuring the temperature inside the aircraft (\textit{air$_{1}$}), we might also want to track the location of the aircraft at the same time. The framework enables the creation of parallel provenance records since each data point is mapped to a list of associated provenance records~(see \cref{sec:generic-provenance-layer}). We can create one provenance record $\textit{prov}_{\textit{temperature}}(\textit{air}_{1})$ representing the latest temperature inside the aircraft, and one provenance record $\textit{prov}_{\textit{location}}(\textit{air}_{1})$ representing the latest location of the aircraft. Hence, we regard \Cref{req:parallel} as fulfilled.

%\subsection{Non-functional Requirements}
%\label{sec:eval-nonfunctional}
%\begin{figure*}
%\centering
%\begin{subfigure}[]{\columnwidth} \centering
%\begin{tikzpicture}[baseline]
%\begin{axis}[
%   	title={},
%   	height=5cm,
%   	width=8cm,
%   	xlabel={Context length},
%   	ylabel={Gas per provenance record},
%%   xmin=0, xmax=100,
%    ymin=0, ymax=2500000,
%   	legend pos=north west,
%   	ymajorgrids=true,
%   	grid style=dashed,
%]
%\addplot [blue, ultra thick] table [x=contextSize, y=consumedGas, col sep=semicolon] {data/create-context.csv};
%\end{axis}
%\end{tikzpicture}
%\caption{Increasing the Length of the Context Parameter}
%\label{fig:creation-parameter-context}
%\end{subfigure}
%%
%\begin{subfigure}[]{\columnwidth} \centering
%\begin{tikzpicture}[baseline]
%\begin{axis}[
%   	title={},
%   	height=5cm,
%   	width=8cm,
%   	xlabel={No. of input provenance records},
%   	ylabel={Gas per provenance record},
%%    xmin=0, xmax=100,
%    ymin=0, ymax=2500000,
%%    xtick={0,20,40,60,80,100},
%%  	 ytick={0,20,40,60,80,100,120},
%   	legend pos=north west,
%   	ymajorgrids=true,
%   	grid style=dashed,
%]
%\addplot [blue, ultra thick] table [x=inputProvenanceSize, y=consumedGas, col sep=semicolon] {data/create-input.csv};
%\end{axis}
%\end{tikzpicture}
%\caption{Increasing the Number of Input Provenance Records}
%\label{fig:creation-parameter-input}
%\end{subfigure}
%\caption{Gas Consumption for Provenance Creation with Respect to an Increasing Input Parameter Size}
%\label{fig:creation-parameter}
%\end{figure*}
\textit{Integrity~(\Cref{req:integrity}).}
The presented framework uses blockchain technology to provide trustless and tamper-proof storage of provenance records for IoT data. Thus, once records have entered the system, the integrity of records depends on the underlying blockchain technology. Ethereum is a public blockchain with around 8250 fully validating nodes securing the network at the time of writing\footnote{26 July 2019 (\url{https://ethernodes.org/network/1})}. Hence, we consider the integrity requirement for data and computations on Ethereum as fulfilled. However, the framework also needs to provide mechanisms to ensure only correct records enter the system in the first place. As a first step, we implement the concept of ownership of data points using tokens to prevent arbitrary clients from creating provenance records. In future work, this concept could be further extended by mechanisms that guarantee the correctness of the records themselves. 

\textit{Availability~(\Cref{req:availability}).}
In public blockchain networks like Ethereum potentially anyone can run a full node. Hence, we consider the availability requirement as fulfilled.

\textit{Privacy~(\Cref{req:privacy}).}
While the privacy requirements of a provenance system largely depend on the concrete use case at hand, in scenarios with sensitive data, such as a health monitoring system, privacy is crucial~\cite{griggs2018healthcare}. However, privacy remains an ongoing challenge in the realm of (public) blockchains, since a blockchain's security relies on data being transparent and verifiable by every participant. A blockchain-based data provenance framework suffers from the same problem. While some propose to use private blockchains in scenarios where privacy is important~\cite{griggs2018healthcare}, this is not sufficient in scenarios which also depend on the system being publicly available, such as global vaccine supply chains.

%We propose to use sidechains as a technique to overcome the blockchain's inherent scalability issues. Sidechains are in principle completely separate blockchains, they are able to choose their own consensus algorithm and set up restrictions on who can join. Therefore, a sidechain can potentially improve privacy, since data that is stored on a sidechain does not necessarily have to be accessible globally.% Instead, it just has to be accessible to the participants of that particular sidechain. However, the focus of this paper is more on the scalability aspects of sidechains. 

\begin{figure}\centering
\begin{tikzpicture}[baseline]
\begin{axis}[
ybar,
height=4cm,
width=8cm,
bar width = 5,
enlargelimits=0.15,
legend style={font=\footnotesize,at={(0.5,1)}, anchor=north, legend columns=-1},%at={(0.5,-0.15)},,legend 
ylabel={Throughput (TPS)},
xlabel={Gas price (Gwei)},
domain=1:10,
samples=3,
symbolic x coords={0.5,1,2,5,10,20},
ymin=0.2, ymax=2,
ymajorgrids=true,
grid style=dashed,
xtick={0.5,1,2,5,10,20},
%nodes near coords,
%nodes near coords align={vertical},
]
\addplot[
	green,
	fill=green!20!white,
	error bars/.cd,
	y dir=both, y explicit,
] coordinates {
%	(0.5,) +- (0,)
	(1,0.572509178) +- (0,0.4784040076) 
	(2,1.298646475) +- (0,0.3797103555)
	(5,1.04018638) +- (0,0.385890031)
	(10,1.126490439) +- (0,0.3331919714) 
	(20,1.225819142) +- (0,0.3868106149)
};
\addplot[
	blue,
	fill=blue!20!white,
	error bars/.cd,
	y dir=both, y explicit,
] coordinates {
%	(0.5,0.2454241296) +- (0,0.2805456421)
	(1,1.608066017) +- (0,0.04224049755)
	(2,1.551054492) +- (0,0.03810158513)
	(5,1.497250152) +- (0,0.09152118733)
	(10,1.681352858) +- (0,0.04618427386) 
	(20,1.544044728) +- (0,0.03367507604)
};
%\addplot[
%	red,
%	fill=red!20!white,
%	error bars/.cd,
%	y dir=both, y explicit,
%] coordinates {
%%	(0.5,0.568245783) +- (0,0.096548615)
%	(1,0.546849549) +- (0,0.138765655)
%	(2,0.524926451) +- (0,0.1215838565)
%	(5,0.611548654) +- (0,0.1289372899)
%	(10,0.5906354155) +- (0,0.08035585268)
%	(20,0.5795378483) +- (0,0.10103727787)
%};
\legend{Ropsten,Rinkeby}
\end{axis}
\end{tikzpicture}
\caption{Avg. Throughput for Creating Provenance Records}
\label{fig:throughput}
\end{figure}
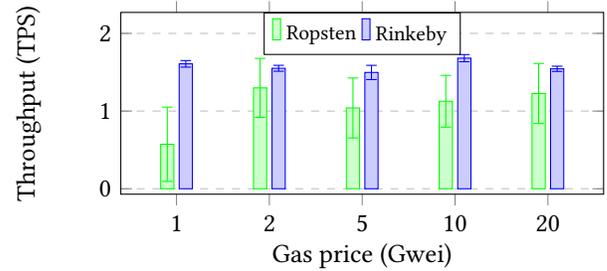

\textit{Scalability~(\Cref{req:scalability}).}
Since IoT scenarios potentially deal with massive amounts of data and frequent data updates, we evaluate the framework's scalability with regard to latency and transaction throughput. Latency and throughput are mostly influenced by two factors, network load~(how many transactions are submitted to the blockchain) and gas price~(how much are the senders of the individual transactions paying as fee). With a higher number of pending transactions (i.e., higher network load), not every transaction can be included within the next block of the blockchain which means higher priced transactions get prioritised since block validators earn more per each included transaction. As the gas price is increased, the average latency converges to the average block time~($\sim$14-30 seconds) of the blockchain, since higher-priced transactions are almost certainly included within the next block. To evaluate throughput, we submit up to 150 transactions to the test network during a 60 second time frame. Afterwards, we divide the number of confirmed transactions with the time frame of 60 seconds to calculate the transactions per second~(TPS). \cref{fig:throughput} shows the average transaction throughput of our experiments. The tradeoff that blockchain-based systems make in terms of scalability and security becomes clear with an average throughput of only 1-1.6~TPS. Increasing the gas cost does not seem to improve throughput beyond a certain point. Notably, throughput was measured from a user perspective, i.e., transactions were submitted from a single account. The maximum global throughput of the Ethereum blockchain is roughly 20~TPS.

To sum up, the framework fulfils all functional requirements defined in \cref{sec:requirements}. By leveraging block\-chain technology, the non-functional requirements of integrity and availability can be fulfilled. Regarding scalability and privacy, the use of blockchain technology poses limitations. Throughput and latency of the framework are constrained, and the inherent transparency of blockchain technology naturally presents a challenge for privacy-sensitive applications.

\section{Related Work}
\label{sec:related}

As mentioned in Section~\ref{sec:intro}, other works~\cite{baracaldo2017securing, polyzos2017blockchain} focus on providing generic blockchain-based data provenance solutions for the IoT. However, these approaches remain conceptual with no provided implementation. In contrast, the framework presented in the work at hand has been fully implemented and is available as open-source software. Other solutions related to blockchain-based data provenance in the realm of the IoT go beyond conceptual approaches~\cite{toyoda2017novel, westerkamp2018blockchain, wu2017distributed, griggs2018healthcare, cebe2018block4forensic}. While the approaches presented in \cite{westerkamp2018blockchain, toyoda2017novel, wu2017distributed} focus on securing provenance information in supply chains, the works in~\cite{griggs2018healthcare} and \cite{cebe2018block4forensic} focus on a blockchain-based patient monitoring system to provide notifications in case of medical emergencies, and a framework that tracks information about vehicles to resolve disputes between drivers, insurance companies and maintenance server providers in case of accidents, respectively. In general, these solutions demonstrate how provenance concepts can be brought onto the blockchain~(e.g., via smart contracts). However in contrast to our work, though IoT technologies are addressed, these solutions are use case-specific and do not provide a generic provenance framework that is required for the heterogeneous nature of the IoT~\cite{olufowobi2017data, li2015internet}.

Provenance in the supply chain has also received attention in the industry. Startups like Provenance\footnote{\url{https://www.provenance.org/}} and Everledger\footnote{\url{https://www.everledger.io/}} focus on securely tracking goods traveling along a supply chain via blockchain technology. However, their sole focus lies on tracking physical items instead of any kind of data within the IoT. Further, works have also focused on blockchain-based data provenance solutions outside the IoT domain, e.g., blockchain-based data provenance for cloud environments to verify the operation history of data in the cloud~\cite{tosh2019data}, or in the scientific field where blockchain technology is used to secure the derivation history of scientific data~\cite{ramachandran2018smartprovenance}.

%Liang et al. present ProvChain~\cite{liang2017provchain}, a blockchain-based data provenance solution for cloud environments. ProvChain provides the ability to record and verify the operation history of data on the cloud. Whenever an operation (read/write) happens on a file in the cloud, the cloud service provider stores a provenance record locally and anchors the hash of the record to a public blockchain. This way, a timestamp proof of that particular operation exists which is verifiable in the future. The information saved on the blockchain is used for the verification of a provenance record, however, it does not provide any provenance information itself. While users can verify that certain operations on a file have occurred, they still depend on a central entity to provide them with a complete set of provenance data for a particular file. 
%
%Secure provenance for scientific data is the focus of~\cite{ramachandran2018smartprovenance}. Here, the authors present~DataProv, a system which securely captures scientific provenance data. DataProv uses the Open Provenance Model standard~\cite{moreau2011open} for modelling provenance information. The system provides an access control mechanism built on smart contracts on the Ethereum blockchain platform to control changes of documents in the cloud while taking the actual verification process of document changes off the chain. This privacy-focused solution looks promising also for IoT-related domains. However, the document-centric provenance model used is problematic for IoT applications~\cite{olufowobi2017data}.

As can be seen by the discussion, to the best of our knowledge, there is no other approach which provides a generic framework for tracking data provenance in the IoT.
\section{Conclusion}
\label{sec:conclusion}

The presented framework provides a data provenance solution which is appropriate for the heterogenous nature of the IoT. By leveraging a generic data model together with a layered architecture of smart contracts, the framework is able to fulfil the functional requirements. By building on blockchain technology, the framework further fulfils the non-functional requirements of integrity and availability but has some limitations regarding scalability and privacy. Multiple current blockchain developments seek to overcome these issues without jeopardising blockchain security, e.g., state channels, zero-knowledge proofs, and sidechains~\cite{eberhardt2017or}. Thus, in future work, we will evaluate in detail to what extent these approaches provide solutions to the privacy and scalability problem in the context of blockchain-enhanced IoT applications, such as our data provenance framework.

\begin{acks}
The work presented in this paper has received funding from
Pantos GmbH within the TAST research project.
\end{acks}
%We would like to thank Christoph Ritzer, Daniel McDonald, Taneli Hukkinen, and Oskar Hladky for their valuable input.
%The work presented in this paper has received funding from Pantos GmbH within the TAST research project.#
%\noindent \hlr{8-10 pages}\\
%\hlr{update access dates}

% TODO MB
%\balance
%\bibliography{refs}
\printbibliography

\end{document}